\documentclass[journal]{IEEEtran}
\usepackage[left=1.43cm,right=1.43cm,top=1.8cm,bottom=4.21cm]{geometry}

\usepackage[algo2e]{algorithm2e} 
\usepackage{amsmath,amssymb,amsmath,amsfonts}
\usepackage{bm}
\usepackage{bbm}
\usepackage{graphicx}
\usepackage{cite}
\usepackage{epstopdf}
\usepackage{gensymb}
\usepackage{color}
\usepackage{lipsum}
\usepackage{float}
\usepackage{epstopdf}
\usepackage[mathcal]{eucal}
\usepackage{subfiles}
\usepackage[T1]{fontenc}
\usepackage{siunitx}

\usepackage{tabulary}
\usepackage{epsfig,graphics,graphicx,amssymb,amstext,amsmath,algorithm,algorithmic,wrapfig,multirow}
\usepackage{array}
\usepackage{adjustbox}
\usepackage[dvipsnames]{xcolor}
\usepackage{cite}
\usepackage{times}
\usepackage{placeins}
\usepackage{textcomp}
\usepackage[font=small]{caption}
\usepackage{flushend}
\usepackage{color, colortbl}
\usepackage{tabularx}
\usepackage{bm}
\usepackage{enumerate}
\usepackage{tikz}
\usepackage{pgfplots}
\usepackage{epstopdf}
\usetikzlibrary{shapes,arrows}
\usepackage[utf8]{inputenc}
\usepackage{mathtools}
\usepackage[utf8]{inputenc}
\usepackage{hyperref}
\hypersetup{
    colorlinks=true,
    linkcolor=black,
    filecolor=magenta,      
    urlcolor=cyan,
    citecolor= blue,
    }
    
\usepackage{xcolor}
\usepackage{tikz}
\usetikzlibrary{chains,arrows,calc,positioning}
\usepackage{amssymb}
\usepackage{pifont}
\usepackage{soul}
\usepackage{breqn} 
\usepackage{booktabs} 
\usepackage{multirow}
\usepackage{enumitem}
\usepackage{tabulary}
\usepackage[normalem]{ulem}
\usepackage{dblfloatfix}
\usepackage{makecell}
\usepackage{lipsum}
\usepackage{amsthm}
\usepackage{comment}
\usepackage{filecontents}
\usepackage{bbm}
\usepackage{glossaries}
\usepackage{graphicx}
\usepackage{subcaption}

\newtheoremstyle{mystyle}
  {}
  {}
  {\itshape}
  {}
  {\bfseries}
  {.}
  { }
  {}
\theoremstyle{mystyle}

\newlength \figwidth
\definecolor{bittersweet}{rgb}{1.0, 0.44, 0.37}
\definecolor{glaucous}{rgb}{0.38, 0.51, 0.71}
\definecolor{gainsboro}{rgb}{0.86, 0.86, 0.86}
\definecolor{babyblueeyes}{rgb}{0.63, 0.79, 0.95}
\definecolor{silver}{rgb}{0.75, 0.75, 0.75}
\definecolor{neoncarrot}{rgb}{1.0, 0.64, 0.26}
\definecolor{Gray}{gray}{0.9}
\definecolor{LightCyan}{rgb}{0.88,1,1}
\definecolor{BackgroundLightBlue}{rgb}{0.97,0.97,1}
\definecolor{BackgroundGray}{gray}{0.98}

\newcommand{\angel}[1]{\noindent{\textsf{[Angel]: {\color{red}#1}}} }

\newcommand{\paul}[1]{\noindent { {{$\blacktriangleright$ 
				{\textsf{[Paul]: {\color{blue}#1}}} $\blacktriangleleft$}}}}

\newcommand{\masoud}[1]{\noindent { {{$\blacktriangleright$ 
				{\textsf{[Masoud]: {\color{magenta}#1}}} $\blacktriangleleft$}}}}

\makeatletter

 \let\oldforeign@language\foreign@language
 \DeclareRobustCommand{\foreign@language}[1]{%
   \lowercase{\oldforeign@language{#1}}}

\def\txr{{\text{u}}}
\def\txt{{\text{b}}}

\def\txT{{\text{T}}}

\def\txF{{\text{F}}}
\def\txi{{\text{i}}}
\def\txv{{\text{v}}}

\def\ba{{\boldsymbol{a}}}

\def\bc{{\boldsymbol{c}}}

\def\bee{{\boldsymbol{e}}}

\def\bg{{\boldsymbol{g}}}

\def\bp{{\boldsymbol{p}}}

\def\bs{{\boldsymbol{s}}}

\def\bv{{\boldsymbol{v}}}

\def\bz{{\boldsymbol{z}}}
\def\b0{{\boldsymbol{0}}}

\def\bA{{\boldsymbol{A}}}

\def\bF{{\boldsymbol{F}}}

\def\bH{{\boldsymbol{H}}}
\def\bI{{\boldsymbol{I}}}

\def\bN{{\boldsymbol{N}}}

\def\bW{{\boldsymbol{W}}}

\def\b{{\mathrm{b}}}

\def\d{{\mathrm{d}}}

\def\nb0{{\mathbf{0}}}
\def\nb1{{\mathbf{1}}}

\def\nbtheta{\boldsymbol{\theta}}

\newacronym{mimo}{MIMO}{multiple-input multiple-output}
\newacronym{simo}{SIMO}{single-input multiple-output}
\newacronym{siso}{SISO}{single-input single-output}
\newacronym{upa}{UPA}{uniform planar array}

\newacronym{mlp}{MLP}{multi-layer perceptron}
\newacronym{bs}{BS}{base station}
\newacronym{ue}{UE}{user equipment}
\newacronym{tdd}{TDD}{time division duplexing}

\newacronym{ode}{ODE}{ordinary differential equation}

\newacronym{osm}{OSM}{OpenStreetMap}
\newacronym{los}{LoS}{line of sight}
\newacronym{nlos}{NLoS}{non-line-of-sight}
\newacronym{aoa}{AoA}{angle of arrival}
\newacronym{csi}{CSI}{channel state information}
\newacronym{csir}{CSIR}{channel state information at the receiver}
\newacronym{aod}{AoD}{angles of departure}

\newacronym{sinr}{SINR}{signal-to-interference-plus-noise ratio}
\newacronym{snr}{SNR}{signal-to-noise ratio}
\newacronym{mmse}{MMSE}{minimum mean squared error}
\newacronym{mse}{MSE}{mean squared error}

\newacronym{awgn}{AWGN}{additive white Gaussian noise}
\newacronym{pas}{PAS}{power angle spectrum}
\newacronym{dsft}{DSFT}{discrete-spatial Fourier transform}
\newacronym{dft}{DFT}{discrete Fourier transform}
\newacronym{pdf}{PDF}{probability density function}

\newacronym{3gpp}{3GPP}{3rd generation partnership project}
\newacronym{umi}{UMi}{urban micro}
\newacronym{uma}{UMa}{urban macro}

\newacronym{svd}{SVD}{singular value decomposition}
\newacronym{iid}{IID}{independent identically distributed}
\newacronym{ao}{AO}{alternative optimization}
\newacronym{se}{SE}{spectral efficiency}

\newacronym{sd}{SD}{schedule deviation}
\newacronym{imcf}{IMCF}{ideal model-consistent flow}

\newacronym{cfmm}{cFMM}{conditional flow matching model}
\newacronym{cddim}{cDDIM}{conditional denoising diffusion implicit model}

\newacronym{film}{FiLM}{feature-wise linear modulation}

\newacronym{gt}{GT}{ground truth}
\newacronym{rt}{RT}{ray-tracing}

\makeatother

\IEEEoverridecommandlockouts
\begin{document}

\bstctlcite{IEEEexample:BSTcontrol}

\title{Site-Specific MIMO Channel Generation via Diffusion and Flow Matching: Fidelity, \\ Efficiency, and Downstream Utility}

\author{
\IEEEauthorblockN{
{Sina Beyraghi, \emph{Graduate Student Member, IEEE}, Masoud Sadeghian, \emph{Graduate Student Member, IEEE},\\ Firdous Bin Ismail, Angel Lozano, \emph{Fellow, IEEE}, Paul Almasan, and Giovanni Geraci, \emph{Senior Member, IEEE}}
\thanks{S.~Beyraghi is with Telefónica Scientific Research and Universitat Pompeu Fabra,
Spain. 
M.~Sadeghian, F.~Bin Ismail, and A.~Lozano are with Universitat Pompeu Fabra, Spain. 
P.~Almasan is with Telefónica Scientific Research, Spain. 
G.~Geraci is with Nokia and Universitat Pompeu Fabra, Spain.}
\thanks{This work was supported in part by the SNS JU Horizon Europe Project under Grant Agreements 101139161 (INSTINCT) and 101192369 (6G-MIRAI), by the Spanish Research Agency through grants PID2021-123999OB-I00, PID2024-156488OB-I00, CEX2021-001195-M, and CNS2023-145384, and by AGAUR.}
}
\thanks{To facilitate reproducibility, we publicly release the complete training and evaluation pipeline, including data pre-processing, model training, channel sampling, metric computation, and downstream task evaluation \cite{our_repo}.}
}

\maketitle

\begin{abstract}

This paper explores the use of generative models to synthesize high-quality, site-specific multiple-input multiple-output (MIMO) channel data, addressing the high cost of the extensive measurement campaigns required to acquire real-world data for AI-native wireless networks. Two location-conditioned generative paradigms are compared: a conditional denoising diffusion implicit model (cDDIM), and a conditional flow matching model (cFMM). Both these models generate MIMO channel matrices conditioned on user coordinates, to preserve the spatial structure of the deployment site. The approaches are evaluated across three dimensions: statistical fidelity (including beam consistency and effective rank), generation efficiency, and utility in downstream tasks such as channel-state information compression and beam alignment. Results across diverse propagation scenarios (28 GHz and 3.5 GHz, both line-of-sight and non-line-of-sight) demonstrate that both models accurately capture site-specific characteristics, even when trained on scarce ground-truth data. Notably, cFMM achieves a quality comparable to cDDIM with roughly an order of magnitude less inference time. Augmenting scarce site-specific datasets with these synthetic channels yields hefty performance gains in downstream physical layer tasks compared to using scarce data alone or stochastic channels.

\end{abstract}



\begin{figure}[t]
    \centering
    \includegraphics[width=0.99\columnwidth]{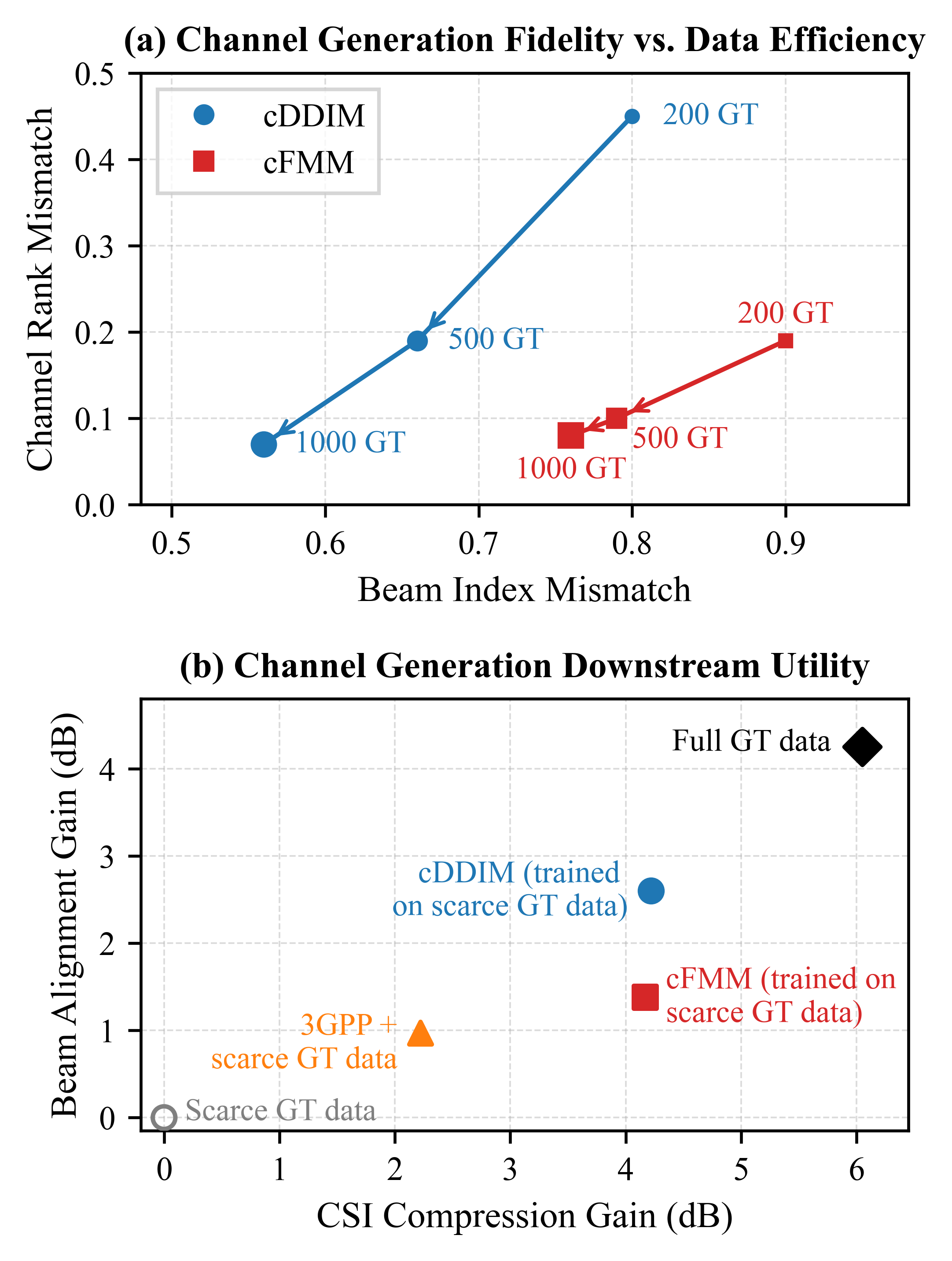}
    \caption{(a) Average mismatch between reference and generated channels at $3.5$ GHz with LoS+NLoS, parameterized by the number of GT training samples.
    The mismatch is measured by the difference between the indices of the dominant beam (horizontal axis) and the Wasserstein distance between the channel ranks (vertical axis). (b) Downstream gains with scarce GT data---200 samples---relative to various alternatives. The gains are   
    for CSI compression at 3.5 GHz with LoS+NLoS (horizontal axis) and beam alignment at 28 GHz with four probing beams (vertical axis). Note: Across all datapoints, cFMM incurs 63× lower sampling latency per generated channel than cDDIM.}
    \label{fig:graphical_summary}
\end{figure}

\section{Introduction}
\label{sec:Sec1}

\subsection{Motivation and Background}

Future wireless systems are envisioned as AI-native, with architectures in which learning is integrated across the radio access network and, in particular, within the physical layer \cite{Reuters_2025_Nvidia_Nokia,KimLeeKim2025,park2026channelgeometrypreservinggenerative,HenZhaAlk2024}.
This requires high-quality, site-specific radio data throughout the models' lifecycle. During training, the data must capture the propagation characteristics of the intended deployment environment; models trained on generic datasets often underperform in site-specific conditions \cite{3gpp_r1-2506243}. Then, upon evaluation, AI-based methods must be benchmarked against strong conventional baselines under site-realistic conditions. Such evaluation is essential to identify the use cases in which AI provides meaningful gains and to quantify the robustness of these gains \cite{oai_10th_anniversary_2024}.

As acquiring high-quality site-specific data through measurement campaigns is time-consuming, costly, and often infeasible at scale, a promising instrument to relieve this bottleneck are generative models that synthesize additional channel realizations from limited site-specific data. Such models can augment measured datasets, enabling the training and evaluation of AI-native radio functions under site-realistic propagation conditions without the need for exhaustive data collection in every environment.

Prior work has explored generative modeling to synthesize wireless channel data, complementing analytical, geometry-based, and stochastic models. Early works and surveys established the potential of generative adversarial networks (GANs) and related schemes to learn channel distributions from observations or simulations \cite{yang2019generative,oshea2019approximating,juhava2023wireless,liu2022channel}. These approaches are attractive because they can capture complex channel statistics without an explicit parametric representation.
%
It was shown in \cite{xiao2022channelgan,xie2023real,lee2024deep} that adversarial learning can synthesize channel samples with statistics resembling those of the training data. Several works further introduced conditioning mechanisms, whereby the generation process is conditioned on transmit--receive antenna coordinates to learn spatially dependent multiple-input multiple-output (MIMO) channel distributions \cite{orekondy2022mimo}. GAN-based channel generation has also been considered for massive MIMO
\cite{euchner2024gan} and for FR3-band propagation across multiple frequencies \cite{hu2024channel}. While these studies demonstrate the promise of GANs for channel synthesis, adversarial training may suffer from instability and mode collapse, and many evaluations focus mainly on selected channel statistics.

A related line of work incorporates scenario awareness, spatial consistency, or temporal structure into generative channel models. For millimeter-wave aerial channels, \cite{xia2022generative} devised a two-stage generative model that first predicts the link state and then generates path losses, delays, and angles. Other works have addressed spatially consistent air-to-ground channel generation from aerial trajectories and received-signal-strength sequences \cite{giuliani2024spatially}, GAN-based generation of air-to-ground multipath parameters conditioned on transceiver location and velocity \cite{tian2024generative}, GAN-based digital-twin channel modeling \cite{zhang2023generative}, space-time predictive channel modeling \cite{li2024gan}, and map-conditioned neural surrogates for link-level pathloss prediction \cite{hehn2023transformer}. These works highlight the value of conditioning generative models on the environmental context, but they often target large-scale quantities or parametric channel descriptions rather than full high-dimensional MIMO channel realizations.

More recently, diffusion models have been introduced for wireless channel modeling and sampling \cite{sengupta2023generative,ChaKumCho2025,JusLeeXin2025}. In \cite{LeeParKim2025}, a conditional diffusion model was applied to high-dimensional user-specific MIMO channel generation, with evaluation in downstream tasks such as channel compression and beam alignment---even if only for sparse millimeter-wave channels with uniform linear arrays, entailing a single angular dimension. 
Compared with GANs, diffusion models provide a stable training framework and can generate diverse samples, but sampling typically requires iterative denoising and may be computationally expensive for high dimensionalities. This motivates comparisons with alternative continuous generative approaches such as flow matching~\cite{lipman2022flow}, that may offer a better tradeoff between fidelity and sampling efficiency.

The evaluation of generative channel models is nontrivial. Recent work has emphasized that generic cost functions may not fully capture wireless physical-layer relevance, and has advocated assessments that reflect both physical consistency and task-level usefulness \cite{baur2024evaluation}. Motivated by this perspective, this paper evaluates site-specific MIMO channel generation along three complementary dimensions: fidelity to the target channel distribution, sampling efficiency, and downstream utility for wireless tasks.
The evaluation extends to planar arrays and to both millimeter-wave and sub-6~GHz channels with rich multipath.


\subsection{Methodology and Contributions}

To the best of our knowledge, this is the first work that evaluates both diffusion and flow-matching for site-specific MIMO channel generation under the aforementioned criteria of fidelity, efficiency, and downstream utility.
Both generative paradigms learn the distribution of the channel matrix conditioned on the \gls{ue} location, allowing synthetic channels to remain tied to the spatial structure of the deployment site: nearby \gls{ue} locations yield channels with consistent angular and spatial characteristics, while locations subject to different propagation conditions produce appropriately distinct channel realizations.

Two paradigms are considered. The first is a \gls{cddim}, which generates channel samples through an iterative reverse denoising process. Diffusion-based generation is attractive because of its ability to model complex high-dimensional distributions, but it typically requires multiple sampling steps at inference time. The second paradigm is a \gls{cfmm}, which learns a deterministic transport field from a latent distribution to the channel distribution. This provides a lower-latency alternative that can generate channels with fewer numerical integration steps. Both paradigms share a common channel representation, conditioning mechanism, and neural backbone, enabling a direct comparison. 

The main contributions of the paper are as follows:
\begin{itemize}

    \item Embodiments of the respective paradigms are developed and compared: a \gls{cddim} model based on iterative denoising, and a \gls{cfmm} model based on deterministic probability transport. Both are designed to generate complex-valued MIMO channel matrices conditioned on \gls{ue} coordinates.

    \item Channel generation fidelity is evaluated using complementary metrics that capture various propagation properties. These include dominant \gls{bs} beam consistency, full beamspace power profile similarity, and effective MIMO channel rank.

    \item The efficiency of the proposed generators is studied from two perspectives: pre-training data efficiency and inference sampling efficiency. This analysis quantifies the trade-off between generation quality, number of available site-specific ground-truth samples, and sampling latency.

    \item The downstream utility of the generated channels is assessed in two representative 
    tasks, namely 
    channel compression and 
    beam alignment. This 
    establishes whether, besides matching statistical channel features,
    synthetic channels improve 
    the learning performance.

\end{itemize}

To facilitate reproducibility and reuse, the code implementing the complete training and evaluation pipeline, including data pre-processing, model training, channel sampling, metric computation, and downstream task evaluation, is freely available \cite{our_repo}.%
%

\subsection{Summary of Results}

A compact visual summary of the main findings is provided in Fig.~\ref{fig:graphical_summary}. 
Multiple insights emanate from the extensive evaluations in the paper, namely:

\begin{itemize}
    \item 
    Both generative models preserve the site-specific spatial structure of the \gls{gt} channels across carrier frequencies and propagation conditions. Even in the more challenging of the considered scenarios
    (at $3.5$~GHz with both \gls{los} and \gls{nlos} propagation), both models retain meaningful agreement in terms of beamspace and effective rank.

    \item 
    \gls{cfmm} provides a consistently more favorable fidelity--latency trade-off than \gls{cddim}, requiring approximately one order of magnitude shorter inference time for comparable channel-generation quality.

    \item 
    Both generators are effective in limited-data regimes. Increasing the number of site-specific training samples consistently reduces the dominant-beam error and its variance. Although \gls{cddim} is slightly more stable for intermediate and large training sets, both 
    already capture the dominant beamspace structure with limited GT data.

    \item 
    Synthetic channels generated by \gls{cddim} and \gls{cfmm} substantially improve downstream channel-state information (CSI) compression compared with using only scarce site-specific data or augmenting it with generic 3GPP stochastic channels. In both LoS and LoS+NLoS settings, the generated-channel augmentation approaches a full $10$k-sample GT reference once a moderate number of real samples is available.

    \item 
    The generated channels also provide clear gains for site-specific beam alignment. At $28$~GHz, models trained with \gls{cddim}- or \gls{cfmm}-generated channels significantly outperform training with 3GPP stochastic channels and scarce GT data. The \gls{cddim}-based beam alignment performs within roughly $1$~dB of training with $10$k real GT channels, while \gls{cfmm} achieves slightly lower but still competitive performance with much faster channel generation.
\end{itemize}

The remainder of the paper is organized as follows. Sec.~\ref{sec:Sec2} introduces the site-specific MIMO channel generation problem, including the system model, conditional generation objective, and dataset summary. Sec.~\ref{sec:Sec3} presents the proposed conditional generative architectures and their training procedures. Sec.~\ref{sec:channel_generation_fidelity} evaluates the fidelity of the generated channels. 
Sec.~\ref{sec:channel_generation_efficiency} studies generation efficiency with respect to training dataset size and inference latency. Sec.~\ref{sec:channel_generation_downstream_utility} evaluates the usefulness of the generated channels for downstream wireless learning tasks. Finally, Sec.~\ref{sec:conclusion} concludes the paper. 

\section{Site-specific MIMO Channel Generation}
\label{sec:Sec2}

\subsection{System Model}


Consider a \gls{bs} equipped with $N_\txt$ antennas 
serving a \gls{ue} located at 
$ \bc = [x,y]^{\txT}$ on a two-dimensional plane, with fixed height $1.5$~m.
The channel between the \gls{bs} and the \gls{ue} is represented by $\bH \in \mathbb{C}^{N_\txr \times N_\txt}$. 
Given $L$ paths, 
\begin{equation}
\bH = \sum_{\ell=1}^{L} \alpha_\ell \ba_\txr(\theta_\ell^\txr , \phi_\ell^\txr ) \ba_\txt^{*}( \theta_\ell^\txt , \phi_\ell^\txt ),
\label{eq:geometric_channel_model}
\end{equation}
where $\alpha_\ell$ is the complex gain of path $\ell$ 
whereas $(\theta_\ell^\txr,\phi_\ell^\txr)$ and $(\theta_\ell^\txt,\phi_\ell^\txt)$ are its elevation and azimuth angles at \gls{ue} and \gls{bs}, respectively.
In turn, $\ba_\txt(\cdot) \in \mathbb{C}^{N_\txt \times 1}$ and $\ba_\txr(\cdot) \in \mathbb{C}^{N_\txr \times 1}$ are the \gls{bs} and \gls{ue} steering vectors, determined by the array geometry.

The propagation parameters $\{\alpha_\ell, \theta_\ell^\txt, \phi_\ell^\txt, \theta_\ell^\txr, \phi_\ell^\txr\}_{\ell=1}^{L}$, as well as $L$ itself, depend on the propagation environment and on $\bc$. 


\begin{figure*}[h]
\centering
\includegraphics[width=0.99\textwidth]{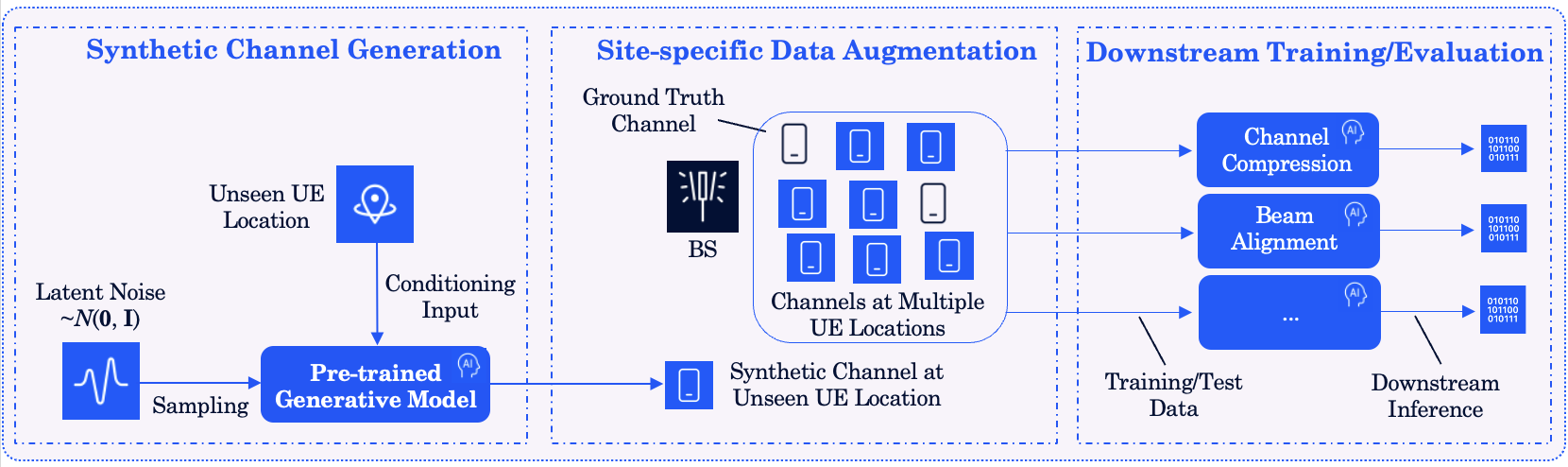}
\caption{Location-conditioned data synthesis. A generative model takes the UE location as conditioning input and performs inference/sampling to synthesize site-specific channels and augment sparse GT data. Augmented datasets are then used to train and/or evaluate downstream models.}
\label{fig:system_model}
\end{figure*}


\subsection{Conditional Channel Generation Objective}

Given a limited set of channel samples and their corresponding \gls{ue} locations, the aim is to train a model that can then generate additional channel realizations conditioned on the \gls{ue} position, as illustrated in Fig.~\ref{fig:system_model}.
%

Let $ \{ (\bH^{(i)},\bc^{(i)}) \}_{i=1}^{M_{\text{GT}}} $
denote the available training dataset of GT channel realizations and corresponding \gls{ue} coordinates, which gives rise to an empirical joint distribution $p_{\text{data}}(\bH,\bc)$. 
The goal is to learn a distribution $p_{\nbtheta}(\bH \mid \bc)$, parameterized by the trainable parameters $\nbtheta$ of the generative model, such that
\begin{equation}
p_{\nbtheta}(\bH \mid \bc)
\approx
p_{\mathrm{data}}(\bH \mid \bc).
\label{eq:learning_objective}
\end{equation}
There are three desirable properties for the learned model:
\begin{enumerate}
    \item 
    \textit{Statistical fidelity:} Generated samples should reproduce the statistical characteristics of the \gls{gt} channels, 
    including angular sparsity, antenna correlations, and spatial consistency across nearby \gls{ue} locations.

    \item
    \textit{Downstream utility:} Generated samples should be useful for downstream communication tasks, i.e., models trained with augmented channel datasets should maintain or improve their performance. This ensures that the generative model captures physically meaningful propagation behavior rather than merely matching low-level statistics.

    \item
    \textit{Training and sampling efficiency:} The generative model should be trainable from a limited number of site-specific channel samples and should enable efficient sampling of new channel realizations once trained.
\end{enumerate}

\subsection{Dataset Summary}
\label{subsec:dataset_summary}

This work employs site-specific \gls{rt} data as a controlled and reproducible proxy for scarce site-specific channel measurements. This choice enables systematic evaluation across carrier frequencies, propagation conditions, training-set sizes, and generative-model sampling budgets. The proposed conditional generation framework is agnostic to how the site-specific channel samples are acquired, and the same training, sampling, and augmentation pipeline can be applied when the available GT data comes from measurement campaigns or measurement-calibrated datasets. 
%
%
GT channels are generated with Sionna \gls{rt}~\cite{aoudia2025sionnart} from a 3D dense urban scene in downtown London, UK. 
The scene covers an area of $200\times200$~$\text{m}^2$ and contains building geometries obtained from \gls{osm}~\cite{haklay2008openstreetmap}.

The \gls{ue} locations are uniformly sampled within a $100$~m radius circular region centered at the \gls{bs}, after which samples inside building meshes are discarded to retain only outdoor \glspl{ue}.
The \gls{bs} is equipped with a $4 \times 8$ \gls{upa}, while each UE features a $2\times2$ UPA. At $28$~GHz, the \gls{bs} height is $8.5$~m; at $3.5$~GHz, it is $25$~m.

For each retained UE position $\bc^{(i)}$, Sionna RT computes the valid propagation paths from the \gls{bs}, including their delays, angles, and complex path coefficients. 
This entails $5\times10^{6}$ rays with up to three bounces, and it accounts for \gls{los} propagation, specular reflections, and diffraction. 
The channel matrix $\bH$ is then constructed as per \eqref{eq:geometric_channel_model}. 

Three dataset variants are assembled, namely $28$~GHz LoS, $3.5$~GHz LoS, and $3.5$~GHz LoS+NLoS. They correspond to the same urban scene, UE sampling procedure, antenna counts, and \gls{rt} 
settings, but differ in carrier frequency and propagation conditions. 
The LoS datasets contain only UE locations with direct BS--UE visibility, whereas the LoS+NLoS dataset contains both LoS and NLoS locations (half each). 
This progression provides a controlled evaluation of conditional channel generation under increasing multipath richness and spatial variability.

For each dataset variant, the UE locations are partitioned into disjoint training, validation, and test sets at the coordinate level. The generative models are trained only on channels from the training coordinates, while fidelity and downstream evaluations are performed on held-out UE coordinates not observed during generator training. Unless otherwise stated, training subsets of size $M_{\rm GT}$ are sampled from the training split, and all reported metrics are computed on the same held-out test split.

\section{Model Architecture and Pre-training}
\label{sec:Sec3}

This section describes the two conditional generative models used to learn $p_\nbtheta(\bH \mid \bc)$.

\subsection{Input Representation}

The spatial structure of MIMO channels is conveniently revealed by a beamspace representation.
The BS \gls{upa} consists of $N_{\txt,x}$ and $N_{\txt,y}$ antennas along the horizontal and vertical dimensions, such that
$
    N_\txt = N_{\txt,x} N_{\txt,y}
$. Let
$
    \bA_{\txt,x} \in \mathbb{C}^{N_{\txt,x} \times N_{\txt,x}}
 $ and $  
    \bA_{\txt,y} \in \mathbb{C}^{N_{\txt,y} \times N_{\txt,y} } 
$ be Fourier matrices, with
$
    \bA_\txt = \bA_{\txt,y}\otimes \bA_{\txt,x}
$
providing a 2D Fourier codebook.
Likewise at the UE side,
$
    \bA_\txr = \bA_{\txr,y} \otimes \bA_{\txr,x}.
$
Then, the beamspace representation is
\begin{equation}
    \bH_\txv = \bA_\txr^* \bH \bA_\txt,
    \label{eq:upa_beamspace_channel}
\end{equation}
where the columns of $\bH_\txv$ index the \gls{bs}-side beams, while its rows index the \gls{ue}-side beams.

The complex beamspace can be further represented as a real-valued tensor  
by assigning the real and imaginary parts of $\bH_\txv$ to two slices along a new component dimension; this tensor serves as input to the generative models conditioned on the \gls{ue} location.

\subsection{Shared Conditional Backbone}
\label{sub:shared_architecture}

Both generative models use the same 
convolutional backbone with residual connections and multi-scale feature aggregation~\cite{ronneberger2015u}. 
The encoder progressively maps the input tensor to lower-resolution feature representations, while the decoder reconstructs the output through a sequence of upsampling stages. 
Skip connections between corresponding encoder and decoder layers preserve fine-grained beamspace information, 
and residual connections within convolutional blocks stabilize training.

The shared architecture ensures that \gls{cddim} and \gls{cfmm} have the same representational capacity. Therefore, differences in their performance can be attributed primarily to the generative objective and sampling procedure, rather than to architectural differences. The neural network contains $15.5\cdot 10^{6}$ parameters.

\subsection{Conditioning Mechanism}

The architecture is conditioned on two variables: $\bc$, and the continuous generative time $t \in [0,1]$.
Rather than directly concatenating them with the input tensor, the conditioning variables are injected into intermediate decoder feature representations through learned modulation layers. 
This allows the shared 
backbone to adapt its decoder features according to both the generation time and the location-dependent propagation conditions, without modifying the underlying convolutional structure.

The conditioning variables are independently embedded using lightweight \glspl{mlp}, 
\begin{align}
\bee_{\text{loc}}^{(k)} & = \phi_{\text{loc}}^{(k)}(\bc) \\
\bee_{\text{time}}^{(k)} & = \phi_{\text{time}}^{(k)}(t)
\end{align}
where $\phi_{\text{loc}}^{(k)}(\cdot)$ and $\phi_{\text{time}}^{(k)}(\cdot)$ denote learnable embedding functions, and $k \in \{1,2\}$ indexes the decoder resolution scale at which the conditioning embeddings are injected.
Two embedding resolutions are employed.
\begin{itemize}
    \item A high-dimensional embedding, 
    applied at the first modulated decoder scale after the bottleneck representation.
    \item A lower-dimensional embedding 
    applied at an intermediate decoder stage.
\end{itemize}
Each embedding is reshaped so that it provides one modulation value per decoder feature dimension, which is then broadcast over the beamspace dimensions of the corresponding decoder feature tensor.

Conditioning is injected into the decoder through the feature-wise modulation operation
\begin{equation}
\bee_{\text{loc}}^{(k)} \odot \bz^{(k)} + \bee_{\text{time}}^{(k)},
\end{equation}
where $\bz^{(k)}$ denotes the decoder feature tensor at scale $k$ while $\odot$ denotes element-wise multiplication. 

Conditioning is applied only in the decoder path after the bottleneck representation has been computed. 
This design choice encourages the encoder to learn generic multi-scale channel representations, while the decoder specializes these representations according to the propagation context and generation timestep. This improves spatial consistency across nearby UE positions and stabilizes training. 

\subsection{Output}

The final layer maps the decoder features back to the same dimensionality as the input real-valued beamspace channel tensor. 
The interpretation of the output depends on the generative model. For \gls{cddim}, the network predicts the noise component. For \gls{cfmm}, it predicts the velocity field.

\subsection{Conditional Diffusion Model}

The \gls{cddim} model entails the complementary 
processes of (i) a forward corruption process that progressively transforms clean samples into Gaussian noise, and (ii) a learned reverse denoising process that reconstructs clean samples from noise \cite{LeeParKim2025}.

During training, a channel sample $\bH_0 \sim p_{\text{data}}(\bH \mid \bc)$ is progressively perturbed according to
\begin{equation}
\bH_t = \alpha_t \bH_0 + \sigma_t \boldsymbol{\epsilon}
\qquad
\boldsymbol{\epsilon} \sim \mathcal{N}(0,\bI)
\qquad t \in [0,1],
\end{equation}
where $(\alpha_t,\sigma_t)$ define a predefined variance schedule controlling the noise level at each timestep.

A neural network $\epsilon_\nbtheta(\bH_t, t,\bc)$ is trained to estimate the noise component added at time $t$ through the objective
\begin{equation}
\mathcal{L}_{\text{cDDIM}} =
\mathbb{E}_{\bH_0,t,\boldsymbol{\epsilon}} \!
\left[
\|
\boldsymbol{\epsilon}
-
\epsilon_\nbtheta(\bH_t,t,\bc)
\|_2^2
\right],
\end{equation}
with $\bc$ being embedded jointly with the temporal encoding, enabling the model to learn a coordinate-dependent denoising function adapted to the UE location.

At inference time, the learned reverse process is used to generate new channel realizations. Starting from a Gaussian noise sample $\bH_1 \sim \mathcal{N}(\boldsymbol{0},\bI)$, the DDIM sampler iteratively removes noise through a sequence of deterministic denoising steps, progressively transforming the latent sample into a clean channel realization
\begin{equation}
\bH_0 \sim p_\nbtheta(\bH \mid \bc ).
\end{equation}
As discussed in Sec.~\ref{sec:channel_generation_efficiency}, the number of denoising steps controls the fidelity--latency trade-off: larger values generally improve reconstruction quality at the expense of increased latency.

\subsection{Conditional Flow Matching Model}


An alternative generative formulation is the state-of-the-art \gls{cfmm}~\cite{lipman2022flow, rombach2022high}. Compared to diffusion-based models, flow matching provides a simpler training objective, 
and typically enables high-quality generation with fewer integration steps. This renders \gls{cfmm} particularly attractive in scenarios where both sample fidelity and low-latency inference are critical.

Rather than a stochastic reverse-time process, \gls{cfmm} learns a continuous-time velocity field that defines a deterministic probability flow between a simple base distribution and the target conditional distribution.
Let $\bH_{0} \sim p_{\text{data}}(\bH \mid \bc )$ be a channel sample and let $\boldsymbol{\epsilon} \sim \mathcal{N}(0,\bI)$ denote a Gaussian noise sample. A continuous interpolation path is defined between the two as
\begin{equation}
\bH_t = (1-t)\bH_0 + t \boldsymbol{\epsilon}
\qquad t \in [0,1],
\label{eq:FMM_base}
\end{equation}
which evolves smoothly from the Gaussian distribution at $t=1$ to the data distribution at $t=0$.

The target velocity field associated with this interpolation path is obtained by differentiating~\eqref{eq:FMM_base} with respect to time,
\begin{align}
u_t(\bH_0,\boldsymbol{\epsilon})
& = \frac{\d \bH_t}{\d t} \nonumber \\
& =
\boldsymbol{\epsilon} - \bH_0.
\end{align}
A neural network
$v_\nbtheta(\bH_t,t, \bc )$
approximates this velocity field over samples drawn from the interpolation path. The training objective is
\begin{equation}
\mathcal{L}_{\mathrm{cFMM}}
=
\mathbb{E}_{\bH_0,\boldsymbol{\epsilon},t}\!
\left[
\left\|
v_\nbtheta(\bH_t,t, \bc)
-
(\boldsymbol{\epsilon}-\bH_0)
\right\|_2^2
\right],
\end{equation}
where $t \sim \mathcal{U}(0,1)$ is sampled uniformly. 

Importantly, $\bc$ is provided as an additional input, allowing the learned velocity field to depend explicitly on the UE position. With that, the model learns a family of coordinate-dependent transport dynamics that map Gaussian noise samples to valid channel realizations consistent with the spatial location.

At inference time, generation starts from a Gaussian sample
$\bH_1 \sim \mathcal{N}(0,\bI)$
and integrates the learned ordinary differential equation 
from $t=1$ down to $t=0$,
\begin{equation}
\frac{\d \bH_t}{\d t}
=
v_\nbtheta(\bH_t,t,\bc),
\end{equation}
progressively transporting the initial noise sample toward the target conditional channel distribution. 

\section{Channel Generation Fidelity}
\label{sec:channel_generation_fidelity}

Given that reference channels are obtained from the dataset described in Sec.~\ref{subsec:dataset_summary} and synthetic channels conditionally produced for the same held-out \gls{ue} locations, it can be established
whether the latter preserve the location-conditioned characteristics of the former.
This evaluation is necessary because,
if the generated channels do not preserve key propagation features, the augmented dataset may introduce systematic biases and degrade downstream performance.

The fidelity is assessed from two complementary perspectives, namely the dominant 
beam directions and the effective rank.
These two metrics are invariant to common phase rotations \cite{LeeParKim2025}, whereby the fidelity evaluations are robust against timing offsets.

%
 
Unless otherwise stated, the results in this section are for 200 site-specific training samples for each conditional generator and a deliberately limited---to curb computing cost and inference time---number of sampling steps: 150 for cDDIM, 10 for cFMM. They should therefore be interpreted as a baseline fidelity assessment under constrained data and inference budgets, with the effect of increasing these budgets studied in Section~\ref{sec:channel_generation_efficiency}.

%


\subsection{Beamspace Power Distribution}
\label{subsec:dominant_beam_index_difference}

Preserving the dominant beam directions is directly relevant to beam selection, as otherwise a suboptimal beamforming direction would be selected. The evaluation is focused on the \gls{bs}-side beamspace structure because the \gls{ue}-side beam response is sensitive to the \gls{ue} orientation~\cite{TR38.901}.

For $q\in\{\mathrm{GT},\mathrm{Gen}\}$, where
$\mathrm{Gen}$ denotes the generated channel, $\bH_\txv^{(q)}$ is the beamspace channel.
The BS beam is indexed by $(i_x,i_y)$, where $i_x \in \{1,\ldots,N_{\txt,x}\}$ and $i_y \in \{1,\ldots,N_{\txt,y}\}$. Letting
\begin{equation}
    \bar{\bH}_\txv^{(q)}
    \in
    \mathbb{C}^{N_\txr \times N_{\txt,x} \times N_{\txt,y}}
\end{equation}
be obtained by mapping the \gls{bs}-side beam dimension of $\bH_\txv^{(q)}$ onto the planar beam grid,
\begin{equation}
p^{(q)}(i_x,i_y)
=
\frac{
\sum_{j=1}^{N_\txr}
\left|
\bar{\bH}_\txv^{(q)}(j,i_x,i_y)
\right|^2
}{
\sum_{i_x=1}^{N_{\txt,x}}
\sum_{i_y=1}^{N_{\txt,y}}
\sum_{j=1}^{N_\txr}
\left|
\bar{\bH}_\txv^{(q)}(j,i_x,i_y)
\right|^2
},
\label{eq:bs_power_distribution_upa}
\end{equation}
represents the fraction of total beamspace power assigned to beam $(i_x,i_y)$.

\subsubsection{Dominant Beam Index}

Let
\begin{equation}
\Big[ i_{x}^{(q)},i_{y}^{(q)} \Big]^\star
=
\arg\max_{(i_x,i_y)}
p^{(q)}(i_x,i_y).
\label{eq:dominant_beam_upa}
\end{equation}
index the strongest beam.
The mismatch between the beam indices of the GT and generated channels can be measured 
\begin{equation}
\mathrm{Beam~Index~Distance}
=
\left\|
\begin{bmatrix}
i_{x}^{\rm(GT)} \\
i_{y}^{\rm (GT)}
\end{bmatrix}^\star
-
\begin{bmatrix}
i_{x}^{\rm (Gen)} \\
i_{y}^{\rm (Gen)}
\end{bmatrix}^\star
\right\|_2^ .
\label{eq:beam_index_distance}
\end{equation}

\begin{figure}[t]
    \centering
\includegraphics[width=1.0\columnwidth]{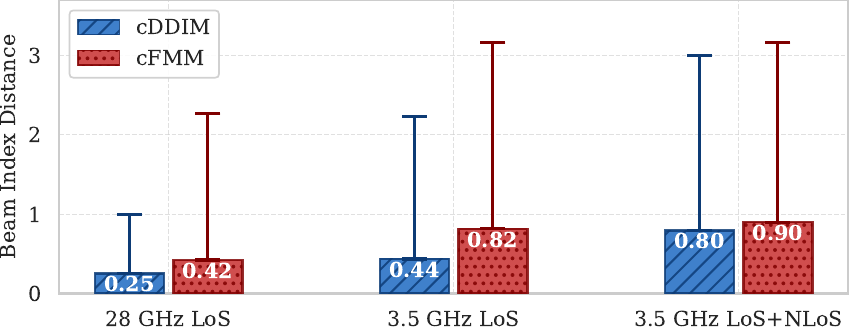}
    \caption{Mean (bar value) and 95th percentile (whisker) of the beam index distance 
    for both generative models trained on 200 samples in three scenarios. 
    }
\label{fig:barplot_final_models_beam_index_distance}
\end{figure}

Fig. \ref{fig:barplot_final_models_beam_index_distance} reports the  beam index distance for \gls{cddim} and \gls{cfmm}. 
The results show a consistent trend, with performance degrading gracefully as the propagation complexity increases.
This indicates that the models' performance is primarily driven by the intrinsic difficulty of the channel distribution, rather than the choice of generative paradigm. At the same time, both models maintain high fidelity even in the most challenging scenario, demonstrating their ability to capture complex, location-dependent propagation effects.

\begin{figure}[t]
    \centering
\includegraphics[width=1.0\columnwidth]{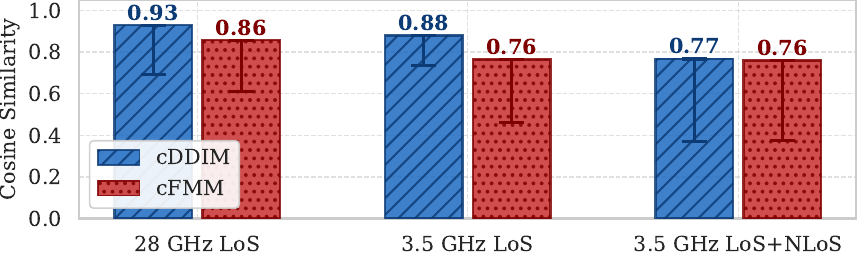}
    \caption{Mean (bar value) and 5th percentile (whisker) of the beamspace power cosine similarity for both generative models trained on 200 samples in three scenarios.}
\label{fig:barplot_final_models_cosine_similarity}
\end{figure}


\subsubsection{Beamspace Power Cosine Similarity}
\label{subsub:cosine_similarity}

The beam index distance assesses whether the generated channel preserves the dominant beam. 
Nevertheless, 
two channels may share the same dominant beam index while differing substantially in their sidelobe levels or angular spread. 
For a broader assessment, a beamspace power cosine similarity metric is introduced that compares the beam power profiles, 
namely
\begin{equation}
\mathrm{Cosine~Similarity} =\frac{ \bp_{\text{GT}}^{\txT}\bp_{\text{Gen}} }{\left\|\bp_{\text{GT}}\right\|\left\|\bp_{\text{Gen}}\right\|},
\label{eq:Cosine_similarity}
\end{equation}
where $\bp_{q}$ is a vector whose $i$th entry equals
\begin{equation}
\bp_q(i)= \frac{ \sum_{j=1}^{N_\txr} \left| \bH_\txv^{(q)}(j,i) \right|^2 }{ \sum_{i=1}^{N_\txt} \sum_{j=1}^{N_\txr} \left| \bH_\txv^{(q)}(j,i) \right|^2 }.
\label{eq:bs_power_distribution_vector}
\end{equation}
Values 
close to one indicate that both channels concentrate power over similar 
beam regions, whereas smaller values reveal discrepancies in the beam power distributions.

The cosine similarity compares the 
power profiles at the same beam indices.
In contrast, the beam index difference focuses only on the location of the dominant beams. 
Thus, the two metrics provide complementary information.

Fig. \ref{fig:barplot_final_models_cosine_similarity} evaluates the beamspace-power cosine similarity 
for the considered scenarios. 
In the $28$~GHz LoS and $3.5$~GHz LoS scenarios, \gls{cddim} better preserves the 
angular power distribution. 
In the more challenging $3.5$~GHz LoS+NLoS setting, both methods achieve comparable average cosine similarities.


\subsection{Effective Channel Rank}
\label{subsec:effective_rank}

To further assess whether the synthetic channels preserve the MIMO structure of the reference dataset, the effective rank of the channel matrix is a suitable metric. 
Let $\sigma_1\geq\sigma_2\geq\cdots\geq\sigma_K$ denote the singular values of the matrix,
where $K=\min(N_\txr,N_\txt)$. The normalized modal powers are then 
\begin{equation}
  p_i =
  \frac{\sigma_i^2}{\sum_{j=1}^{K}\sigma_j^2}
  \qquad i=1,\ldots,K
  \label{eq:effective_rank_prob}
\end{equation}
and one possible manner in which the effective rank can be computed is~\cite{Effective_Rank}
\begin{equation}
  r
  =
  \exp\!\left(
  -\sum_{i=1}^{K} p_i \ln p_i
  \right),
  \label{eq:effective_rank}
\end{equation}
which satisfies $1\leq r\leq K$. A value close to one indicates that most of the
channel energy is concentrated in a single dominant spatial mode, whereas
larger values indicate a more balanced energy distribution across multiple
spatial modes.

The effective-rank distributions of the GT and generated channels are compared by means of the Wasserstein distance. This distance measures the minimum amount of distributional mass that must be transported, weighted by the transport distance, to transform one distribution into the other; it is nonnegative and equals zero only when the two distributions coincide.

\begin{figure*}[!t]
    \centering

    \includegraphics[width=0.98\textwidth]{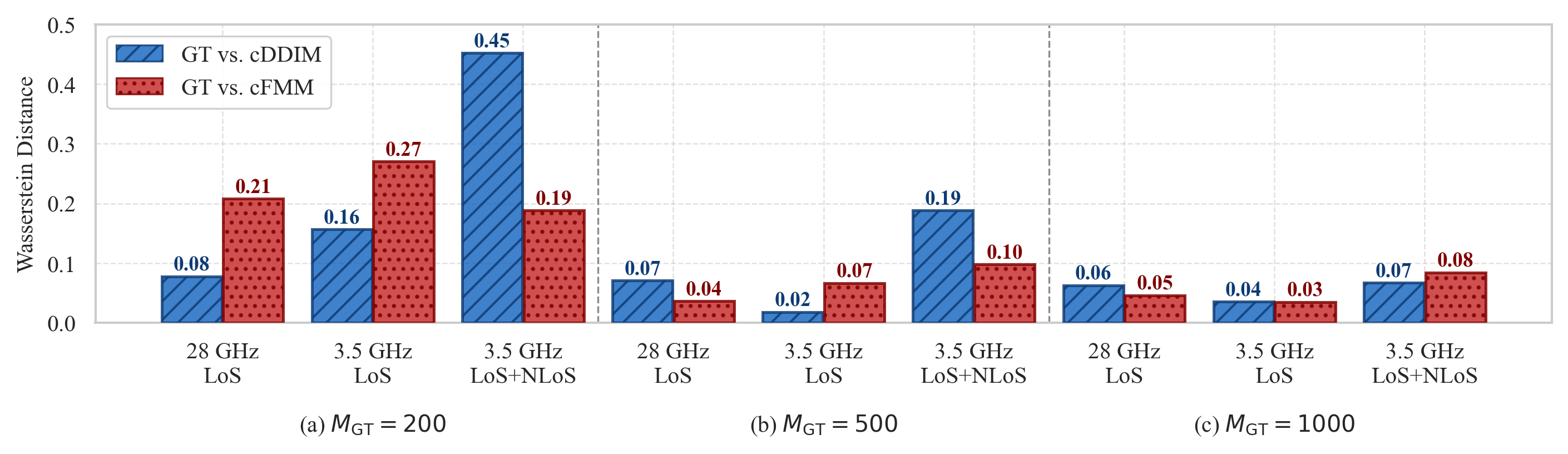}

    \caption{Wasserstein distance 
    between the effective-rank distributions of the GT and the generated channels as a function of the number of GT samples in the training set, for different scenarios. }
    \label{fig:wasserstein_distance_effective_rank}

    \setcounter{subfigure}{0}
    \refstepcounter{subfigure}
    \label{fig:wasserstein_distance_effective_rank_200}

    \refstepcounter{subfigure}
    \label{fig:wasserstein_distance_effective_rank_500}

    \refstepcounter{subfigure}
    \label{fig:wasserstein_distance_effective_rank_1000}

\end{figure*}

Fig.~\ref{fig:wasserstein_distance_effective_rank_200} reports the Wasserstein distance between the effective-rank distributions of the GT channels and those generated by cDDIM and cFMM, with only $M_{\text{GT}}=200$ site-specific training samples. cDDIM achieves significantly smaller distances than cFMM in the LoS-only scenarios, suggesting that cDDIM more accurately preserves the low-rank structure of the LoS reference channels under this condition.
Interestingly, in the mixed LoS+NLoS case at $3.5$~GHz, the trend is reversed: cFMM achieves a lower distance than cDDIM. This suggests that cFMM better captures the broader effective-rank distribution induced by the richer multipath structure of the mixed propagation setting.

\section{Channel Generation Efficiency}
\label{sec:channel_generation_efficiency}

This section evaluates the efficiency of the proposed channel generators along two dimensions.
First, pre-training data efficiency, meaning how the generated channel fidelity scales with the number of site-specific training samples. 
Second, inference sampling efficiency, which measures the fidelity--latency trade-off as the number of sampling steps changes.


\subsection{Pre-training Data Efficiency}
\label{subsec:pretraining_data_efficiency}

For this assessment, \gls{cddim} employs $150$ denoising steps while \gls{cfmm} applies $10$ solver steps. (The effect of varying the numbers of inference steps is studied in Sec.~\ref{subsec:inference_sampling_efficiency}.)
Both are trained with progressively larger subsets of the dataset for the $3.5$\,GHz LoS scenario. Precisely, the number of training samples varies from $M_{\text{GT}}=200$ to $10000$. 

\begin{figure}[t]
    \centering
\includegraphics[width=0.99\columnwidth]{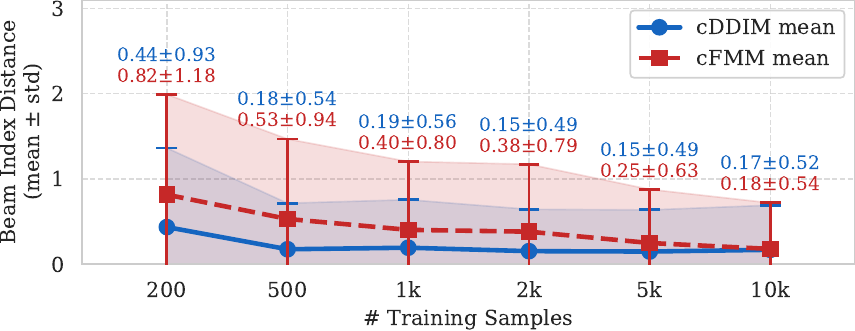}
    \caption{Mean and standard deviation of the beam index difference, 
    trained on varying number of samples for the 3.5 GHz LoS setting. 
    }
\label{fig:training_splits_los}
\end{figure}

Fig. \ref{fig:training_splits_los} presents the beam index distance as a function of the training dataset size. As one would expect, increasing the amount of training data consistently reduces the mean and variance of the beam index error.
Both models exhibit relatively robust behavior even in low-data regimes. A few hundred training samples suffice to capture the dominant beamspace structure. 

\begin{figure}[t]
    \centering
\includegraphics[width=0.99\columnwidth]{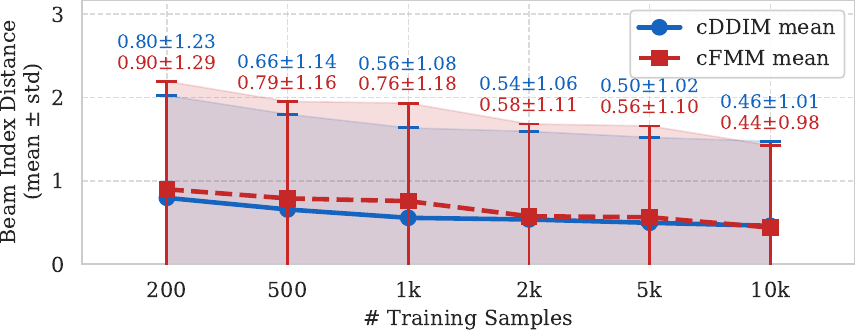}
    \caption{Mean and standard deviation of the beam index distance, trained on varying number of samples for the 3.5 GHz LoS+NLoS setting.}
\label{fig:training_splits_nlos_los}
\end{figure}

For Fig. \ref{fig:training_splits_nlos_los}, the models face the more challenging $3.5$~GHz LoS+NLoS scenario. The higher intricacy leads to higher beam index variability and a broader error spread, yet both models continue to progressively learn the more complex angular propagation patterns associated with NLoS environments. 

Revisiting Fig.~\ref{fig:wasserstein_distance_effective_rank}, it
can be appreciated how increasing the number of training samples from $M_{\text{GT}}=200$ to $1000$ dramatically reduces the Wasserstein distance between the effective-rank distributions of the GT and the generated channels for both cDDIM and cFMM.

\subsection{Inference Sampling Efficiency}
\label{subsec:inference_sampling_efficiency}

The trade-off between generation fidelity and computational cost is gauged by varying the number of sampling steps during inference. For \gls{cddim}, this corresponds to the number of reverse denoising iterations whereas, for \gls{cfmm}, it corresponds to the number of numerical integration steps employed to solve the probability-flow \gls{ode}.
This evaluates how efficiently each generative paradigm transports samples from the latent Gaussian space to the conditional channel distribution.

To this end, multiple instances of each model are evaluated with different discretization granularities.
In particular, four \gls{cddim} models with $\{10, 50, 150, 200\}$ denoising steps and four \gls{cfmm} models with $\{5, 10, 15, 50\}$ solver steps are considered. These configurations span a wide range of computational budgets, from low latency to high fidelity. The quality of the generated channels is evaluated using beamspace power cosine similarity (recall Sec.~\ref{subsub:cosine_similarity}).

\begin{figure}[t]
    \centering
\includegraphics[width=0.99\linewidth]{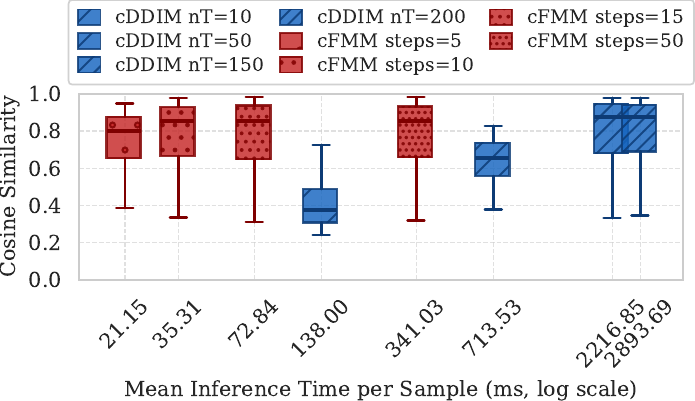}
    \caption{Beamspace power cosine similarity vs. mean inference time for different sampling configurations on the 3.5 GHz LoS+NLoS scenario. The central marker denotes the median cosine similarity, the box outlines the interquartile range (IQR: 25th–75th percentiles) and whiskers extend to \(1.5 \times \text{IQR}\).} 
\label{fig:ablation_inference_time}
\end{figure}

Fig.~\ref{fig:ablation_inference_time} 
shows the fidelity--latency trade-off for the considered sampling configurations on the 3.5 GHz LoS+NLoS scenario. The latency is measured in terms of inference time and several trends can be observed:
\begin{itemize}
    \item Increasing the number of sampling steps improves the reconstruction fidelity for both generative paradigms. For \gls{cddim} especially, it increases very substantially when moving from $10$ to $150$ iterations, evidencing how additional denoising iterations progressively refine the generated beamspace structure. 
    
    \item \gls{cfmm} achieves high cosine similarity with substantially fewer inference steps; $10$--$15$ \gls{ode} steps suffice to attain mean cosine similarities close to $0.9$, and with substantially lower inference time than high-step diffusion configurations. In contrast, \gls{cddim} requires a larger number of denoising iterations to reach a comparable level of fidelity.
    
    \item The figure highlights the fundamentally different scaling behavior of both approaches. 
    \gls{cddim} refines generated channels through successive denoising steps, so additional iterations gradually improve the beamspace structure.
    Conversely, \gls{cfmm} learns a direct deterministic transport field between the latent and data distributions, allowing accurate generation with significantly fewer integration steps.
\end{itemize}

Beyond a point, additional steps yield diminishing returns for both methods. For example, the improvement from $150$ to $200$ iterations in \gls{cddim}, or from $15$ to $50$ steps in \gls{cfmm}, produces only marginal gains in cosine similarity while substantially heightening the computational cost. This suggests that the dominant channel structure is already captured at moderate discretization levels.
Overall, \gls{cfmm} is seen to provide a more favorable fidelity--latency trade-off for conditional channel generation. The learned velocity field enables efficient transport toward realistic beamspace channel distributions using only a small number of deterministic integration steps, making flow matching particularly attractive for low-latency channel generation.

\section{Downstream Utility} 
\label{sec:channel_generation_downstream_utility}

To calibrate the practical usefulness of the proposed channel augmentation framework, two representative downstream learning tasks are entertained, namely CSI compression and site-specific beam alignment. For both downstream tasks, training and augmentation use only the training-coordinate pool, while performance is reported on held-out channels from disjoint test UE locations.


\subsection{CSI Compression}

The objective here is to reduce the feedback overhead while preserving the information required for accurate channel reconstruction. The downlink is considered, with the normalized mean-square error (NMSE) on the CSI as a performance measure. 
The scheme of choice  is CRNet, a deep-learning-based CSI compression architecture originally proposed for multi-resolution CSI reconstruction in massive MIMO \cite{CRNet}. 
To isolate the performance of the CSI compression itself, perfect downlink channel estimation and ideal uplink feedback are considered.

The input downlink channel matrix is first transformed into the beamspace domain by applying a 2D Fourier transform, as described in Sec.~\ref{subsec:dominant_beam_index_difference}. 
The resulting beamspace representation is then passed through the compressing encoder network; then, at the receiver side, the decoder network reconstructs the beamspace representation. Finally, an inverse 2D Fourier transformation maps it back to the antenna domain.
The reconstruction quality is evaluated in the beamspace domain via
\begin{equation}
\mathrm{NMSE}
=
\mathbb{E} \! \left[
\frac{\left\|\hat{\bm{H}}_{\txv}-\bm{H}_{\txv}\right\|_{\txF}^{2}}
{\left\|\bm{H}_{\txv}\right\|_{\txF}^{2}}
\right],
\label{eq:nmse_crnet}
\end{equation}
where $\bm{H}_{\txv}$ and $\hat{\bm{H}}_{\txv}$ are the original and reconstructed
beamspace matrices, respectively, and $\|\cdot\|_{\txF}$ is the Frobenius norm.

Since CRNet training can lead to different local solutions due to random initialization and stochastic gradient-based optimization, each experiment is repeated four times; the model with the lowest validation NMSE is selected. 

\begin{figure}[!t]
\centering

\includegraphics[width=\columnwidth]{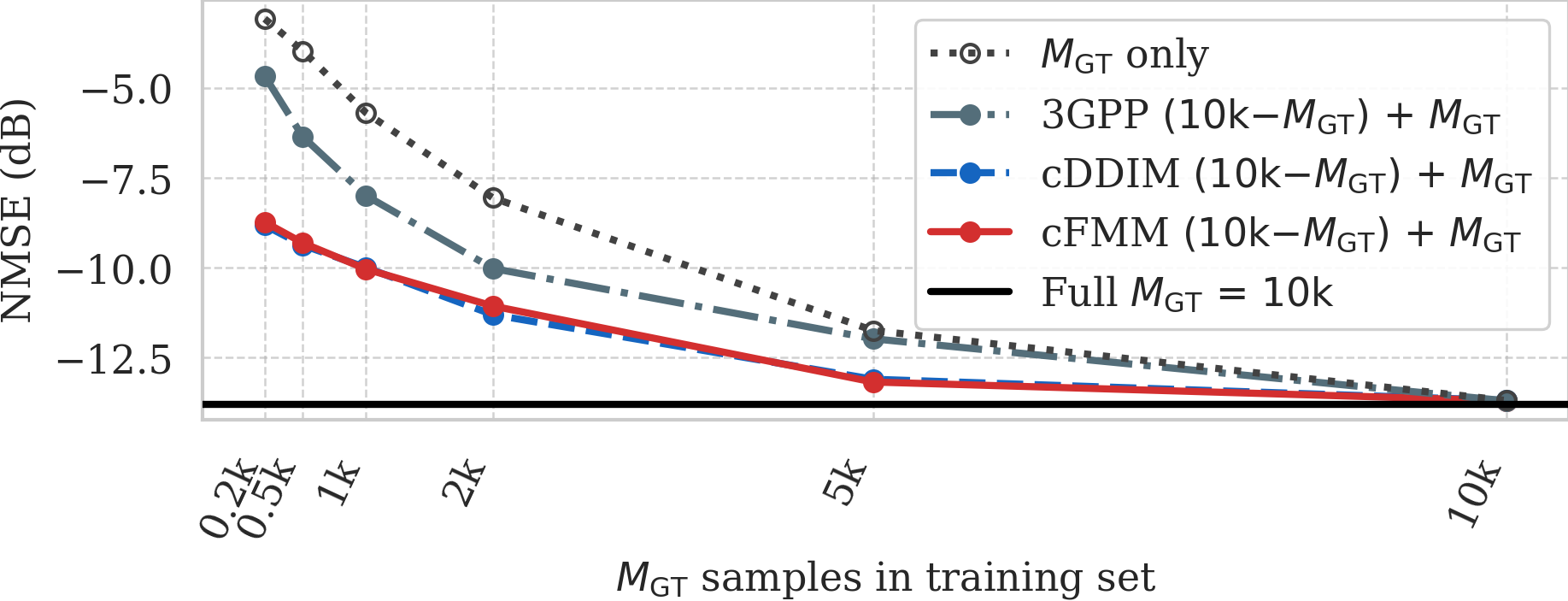}
\vspace{-1mm}

{\footnotesize (a) 3.5\,GHz LoS.}
\vspace{2mm}

\includegraphics[width=\columnwidth]{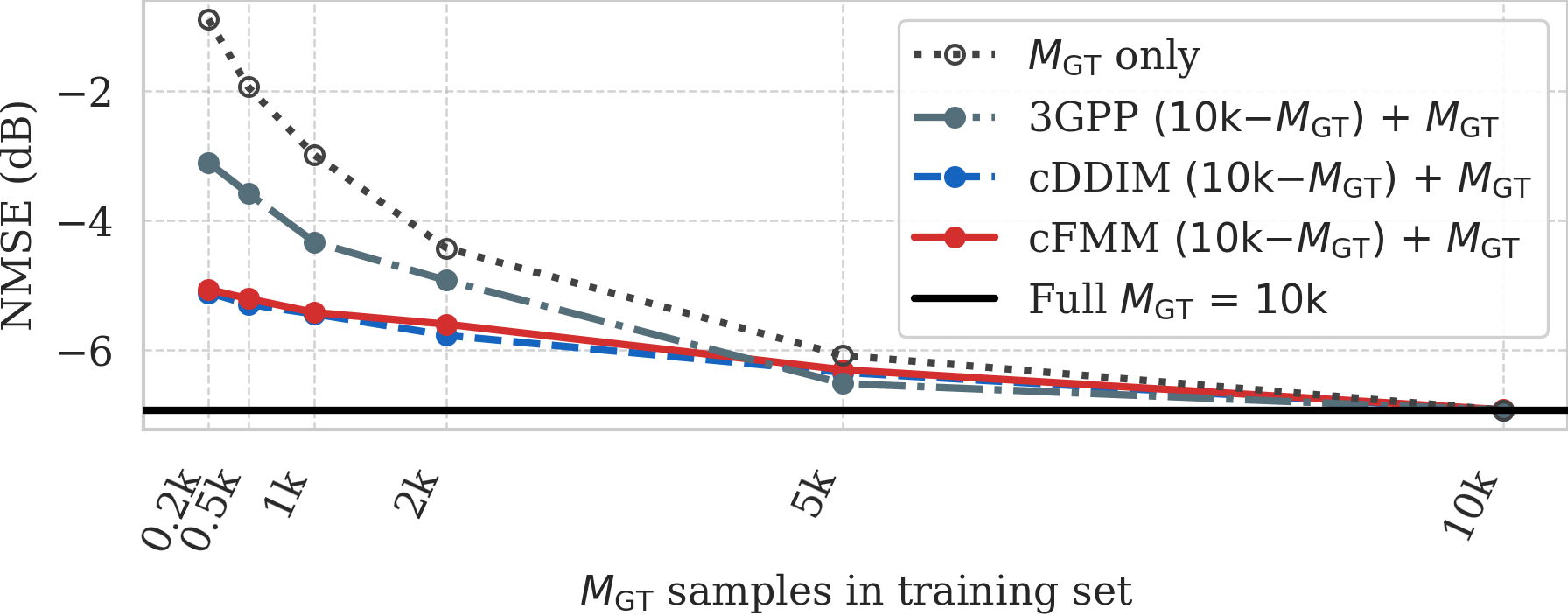}
\vspace{-1mm}

{\footnotesize (b) 3.5\,GHz LoS+NLoS.}

\caption{CSI compression performance in terms of NMSE as a function of the number of GT samples in the training set. 
}
\label{fig:csi_compression_crnet_los_nlos}
\end{figure}


The encoder uses a multi-resolution convolutional design with two parallel 
branches: one applies consecutive convolutional and batch-normalization 
layers with kernels of size $3\times3$, $1\times9$, and $3\times1$; the other one uses a single $3\times3$ convolution. The extracted features are 
concatenated, fused by a $1\times1$ convolution, flattened, and compressed through 
a fully connected layer. 
At the decoder, the latent vector is expanded through a fully connected 
layer and reshaped into the original tensor format. A $5\times5$ convolution is 
applied to obtain an initial reconstruction. 
This is refined by two cascaded CRBlocks, each combining 
multi-kernel convolutional paths with residual connections.
The output is finally passed 
through a sigmoid activation. 

The 3.5~GHz LoS and 3.5~GHz LoS+NLoS scenarios are considered, with
$\{200, 500, 1\mathrm{k}, 2\mathrm{k}, 5\mathrm{k}, 10\mathrm{k}\}$.
For the augmentation-based evaluations, the training set consists of $M_{\text{GT}}$ samples and $10\mathrm{k}-M_{\text{GT}}$
additional generated samples.
Three augmentation sources are contrasted: cDDIM-generated, cFMM-generated, and 3GPP stochastic channel samples generated under the same conditions. Included as benchmarks are the limited number of GT samples without augmentation, and the full set of $10\mathrm{k}$ GT samples.

Fig.~\ref{fig:csi_compression_crnet_los_nlos}a corresponds to the LoS scenario. The benchmark having $10\mathrm{k}$ GT samples achieves NMSE$=-13.8$~dB while, with only $200$ and $500$ GT samples, that rises to $-3.1$~dB and $-4$~dB, respectively. Augmenting these limited datasets with generated channels brings the NMSE back down; with only $M_{\text{GT}}=200$, cDDIM and cFMM respectively achieve $-8.8$~dB and $-8.7$~dB. 
At the same time, the comparison with 3GPP-based augmentation highlights the importance of site specificity. Although produced under the same conditions, stochastic 3GPP channels yield a smaller improvement than the generative models in the important low-data regime. At $M_{\text{GT}}=200$, 3GPP augmentation attains $-4.66$~dB, notably worse than both cDDIM and cFMM.

Fig.~\ref{fig:csi_compression_crnet_los_nlos}b reports results for the more challenging LoS+NLoS scenario. With 10k GT samples, NMSE$=-7$~dB, which rises to
$-0.9$~dB and $-2$~dB for $M_{\text{GT}}=200$ and $500$, respectively. The generative augmentation methods bring that back down; for $M_{\text{GT}}=200$, both cDDIM and cFMM achieve $-5.1$~dB.
3GPP augmentation yields inferior improvements, achieving only $-3.1$~dB for $M_{\mathrm{GT}}=200$. 

Altogether, the results evince that the generated channels provide useful training samples for CSI compression and can significantly reduce the amount of GT data.

\subsection{Beam Alignment} 

The goal here is not to reconstruct the channel itself, but to determine beamforming directions that maximize the received signal power \cite{BAE}.
The 28~GHz LoS setting is considered, as beamforming  acquires growing importance at higher frequencies.

\begin{figure}[!t]
\centering
\includegraphics[width=\columnwidth]{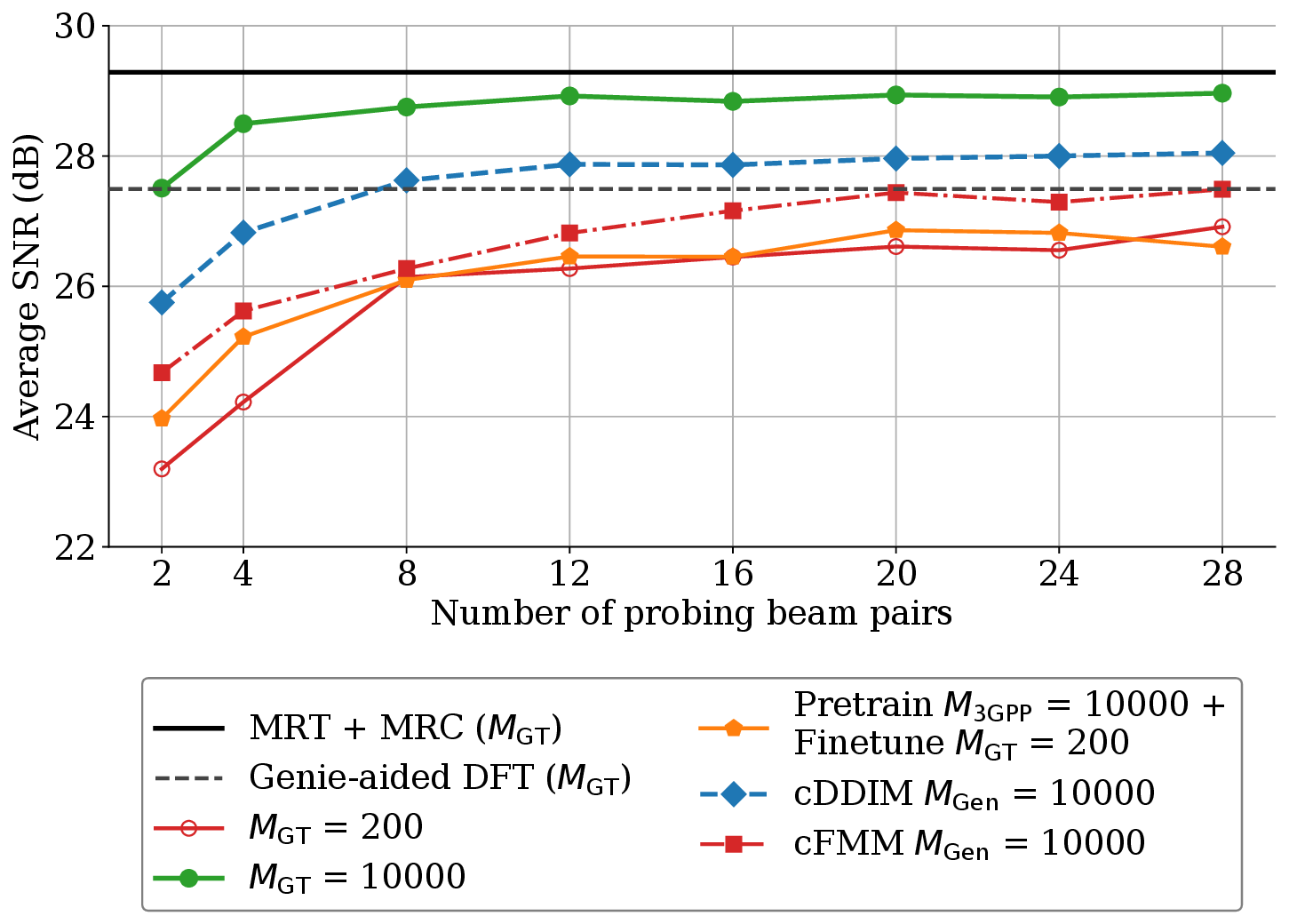}
\caption{Average SNR vs. number of probing beam pairs for the $28$~GHz LoS scenario, comparing downstream evaluation with GT, cDDIM, cFMM, and 3GPP channel data.
}
\label{fig:los_28ghz_sionnart}
\end{figure}

A limited set of probing beams acquires compressed channel observations, which are then processed by a neural network to infer the transmit and receive beamformers.
With the transmitter applying a probing precoder matrix $\bm{F}$ and the receiver applying a combining matrix $\bm{W}$, the observation is
\begin{equation}
\bm{Y} = \sqrt{P}\,\bm{W}^* \bm{H}\bm{F} \operatorname{diag}(\bs) + \bm{W}^* \bm{N},
\end{equation}
where $P$ is the transmit power, $\bs$ is a vector of probing symbols, and $\bN$ is IID complex Gaussian noise with power $\sigma^2 = -81$~dBm, corresponding to a power spectral density of $-161$~dBm/Hz over a bandwidth of $100$~MHz.
From these observations, one can construct the feature vector
\begin{equation}
\bg = \left[ |[\mathrm{diag}(\bm{Y})]_1|^2,\ldots,|[\mathrm{diag}(\bm{Y})]_{N_{\text{probe}}}|^2 \right]^\txT,
\end{equation}
where $N_{\text{probe}}$ is the number of probing beam pairs, with each pair corresponding to a transmit beam from $\bm{F}$ and a receive beam from $\bm{W}$. This vector contains the received powers associated with the probing stage, and serves as input to the neural network.

The beam aligner maps $\bg$ to a transmit beam $\bv_\txt$ and a receive beam $\bv_\txr$. Precisely, a trainable probing stage extracts low-dimensional information, and a beam synthesizer network transforms it into a beam pair. Thus, the learning problem can be interpreted as a mapping from compressed site-specific channel observations to beam decisions. In the implementation herein, the additional initial access term considered in \cite{BAE} is omitted, as it depends on large-scale channel features that are not preserved by cDDIM and cFMM.
Instead,
the generated channels are normalized to enforce a constant Frobenius norm throughout the beam alignment process.

The adopted architecture contains two types of neural components \cite{BAE}.
The first is a complex-valued probing block that learns $\bF$ and $\bW$, and that is parameterized by trainable complex weights. 
The second consists of separate transmit and receive beam synthesizers that process the measured features and generate the final beamforming vectors; both synthesizers rely on fully connected layers, ReLU activation functions, and batch normalization, and their outputs define the real and imaginary parts of the  beamforming vectors used for alignment.

This downstream task is especially suitable to evaluate synthetic channel augmentation. If the generated channels preserve the dominant structure, then the beam aligner trained with such data should learn probing and beam-selection strategies that generalize well to unseen UEs. Conversely, if the synthetic samples fail to capture the spatial signatures of the true site-specific channels, the beam alignment accuracy and the final beamforming gain are bound to deteriorate.

The evaluation metric is the average signal-to-noise ratio (SNR) achieved by the synthesized beams over the test set,
\begin{equation}
\mathrm{SNR}
=
\frac{1}{N_{\text{test}}}
\sum_{i=1}^{N_{\text{test}}}
\frac{P\left|\bv_{\txr,i}^* \bm{H}_i \bv_{\txt,i}\right|^2}{\sigma^2},
\end{equation}
where $N_{\mathrm{test}}$ is the number of test samples.
%
The beam aligner is trained using the datasets produced by the considered augmentation strategies and then evaluated on held-out test channels. As beam alignment depends strongly on the angular and spatial consistency of the channel samples, it provides a meaningful benchmark for judging the quality of the generated data.

Fig.~\ref{fig:los_28ghz_sionnart} illustrates the performance for the 28~GHz LoS dataset as a function of the number of probing beam pairs.
%
The baselines include the maximum-ratio transmission and maximum-ratio combining (MRT+MRC) upper bound 
as well as the genie-aided Fourier method, which selects the best beam pair from the Fourier codebook using the test channel. Across the whole range of $N_{\mathrm{probe}}$, the models trained with \gls{cddim}- and \gls{cfmm}-generated channels consistently outperform both (i) training with scarce GT samples, and (ii) pre-training on stochastic 3GPP data followed by fine-tuning on scarce GT samples. Between the two generative approaches, \gls{cddim} exhibits the strongest performance and approaches the MRT+MRC upper bound, while \gls{cfmm} remains competitive and provides clear gains over the baselines.

These results confirm that the generative models preserve the spatial structure required for effective beam alignment. While cDDIM provides the best performance, cFMM achieves slightly lower but still competitive performance with faster generation. Overall, the comparison confirms that simply increasing the training set with generic stochastic channels is not sufficient; the augmented data must reflect the site-specific channel behaviors to improve beam-alignment performance.


\section{Conclusion}
\label{sec:conclusion}

This paper investigated site-specific MIMO channel generation by means of conditional diffusion and flow-matching models. Channel augmentation was formulated via the learning of the conditional distribution of the channel matrix given the UE location, which enables generated channels to reflect location-dependent propagation effects. 

A unified comparison between cDDIM and cFMM was conducted, 
with both sharing a common channel representation, conditioning mechanism, and neural backbone.
The two approaches satisfyingly reproduce the site-specific structure of the target channel distribution across carrier frequencies and propagation conditions. In particular, the generated channels preserved beamspace power profiles and effective-rank distributions, confirming that the models learn physically meaningful MIMO characteristics rather than only marginal statistics.
The proposed generators were shown to be effective even if only limited site-specific training data are available. Increasing the number of training samples consistently improved generation accuracy and reduced the error variability. Although cDDIM exhibited slightly higher stability for intermediate and larger training sets, both methods were able to capture the dominant beamspace structure from limited data. 

The comparison also revealed a clear difference in generation efficiency. While cDDIM achieved high-fidelity samples after a sufficiently large number of denoising steps, cFMM reached comparable beamspace fidelity with substantially fewer numerical integration steps and approximately one order of magnitude lower sampling latency. This indicates that flow matching is a particularly attractive candidate when fast site-specific channel synthesis is required.

The generated channels were validated through downstream learning tasks. In CSI compression, augmentation with cDDIM- or cFMM-generated channels substantially outperformed both scarce real-data-only training and augmentation with 3GPP stochastic channels. In site-specific beam alignment, the generated-channel augmentation led to significant SNR gains, with cDDIM approaching the performance obtained by training on the full GT dataset while cFMM offered competitive performance with much faster channel generation. These findings confirm that the proposed generative models are useful for training AI models for the wireless physical layer.
\bibliographystyle{IEEEtran}
\bibliography{journalAbbreviations, main}

\end{document}